\documentclass[aps,prl,twocolumn,superscriptaddress,preprintnumbers,%
               showpacs,nofootinbib]{revtex4}
\newcommand{\PRE}[1]{}       % Use if journal style
\usepackage{amsmath}
\usepackage{epsf}
\usepackage{pstricks}

\begin{document}

\preprint{ANL-HEP-PR-04-19, EFI-04-06}

\title{
\PRE{\vspace*{1.5in}}
Warped Unification, Proton Stability and Dark Matter
\PRE{\vspace*{0.3in}}
}

\author{Kaustubh Agashe}
\affiliation{Department of Physics and Astronomy, John Hopkins University, 3400
North Charles St., Baltimore, MD 21218-2686, USA
\PRE{\vspace*{.1in}}
}
\author{ G\'eraldine Servant}
\affiliation{High Energy Physics Division, Argonne National Laboratory, Argonne, IL 60439,
USA
\PRE{\vspace*{.1in}}
}
\affiliation{Enrico Fermi Institute, University of Chicago, Chicago, 
IL 60637, USA
\PRE{\vspace*{.1in}}
}
\PRE{\vspace*{.1in}}
\affiliation{Service de Physique Th\'eorique, CEA Saclay, F91191 Gif--sur--Yvette, France
\PRE{\vspace*{.1in}}
}

%\date{July 7, 2002}

\begin{abstract}
\PRE{\vspace*{.1in}} 
Many extensions of the Standard Model have to face the problem of
new unsuppressed baryon-number violating interactions.
In supersymmetry, the simplest way to solve this
problem is to assume R-parity conservation. As a result,
the lightest  supersymmetric particle becomes stable and a well-motivated 
 dark matter candidate.
In this paper, we show that solving the problem of baryon number violation
in non supersymmetric grand unified theories (GUT's) in warped 
higher-dimensional spacetime can
lead to a stable Kaluza--Klein particle.
This exotic particle has gauge quantum numbers of a right-handed 
neutrino, but carries
fractional baryon-number and is related to the top quark within the 
higher-dimensional GUT. A
combination of baryon-number and $SU(3)$ color ensures its stability.
Its relic density can easily be of the right value
for masses in the 10 GeV--few TeV range. An exciting
aspect of these models is that the entire parameter space will
 be tested at near future dark matter direct detection experiments.
Other exotic GUT partners of the top quark are also light and can be
produced at high energy colliders with distinctive signatures.

\end{abstract}

%12.60.-i   Models beyond the standard model
%11.10.Kk   Field theories in dimensions other than four
%95.35.+d   Dark matter

\maketitle

%\paragraph{Introduction}
One of the most interesting aspects of the dark matter puzzle 
(that $\sim 80 \%$ of the matter in the universe is nonbaryonic 
and of yet-unknown nature) is that it is likely to be related to 
new physics at the TeV scale. Indeed, particles with weak scale size 
interactions and a mass at the electroweak breaking scale (WIMPs) are 
typically predicted to have the good relic density today to 
account for dark matter, provided that they are stable.
The hope is that through the confrontation of collider experiments,
table-top direct searches, neutrino telescope and other cosmic
ray detector experiments, dark matter (DM) will soon reveal itself 
 and with it the first pieces of evidence for new physics at 
 the electroweak scale.
 The favorite DM candidate to date is the Lightest 
 Supersymmetric Particle (LSP) in supersymmetric extensions of the 
 standard model (SM)
 with conserved R-parity. R-parity is not imposed just for the purpose 
 of having a DM candidate but as the simplest solution to 
 baryon number conservation in the supersymmetric theory. As
  a very appreciated spin-off, one gains a stable DM candidate.

Lately, alternative models for physics beyond the SM that make use of 
extra dimensions rather than supersymmetry
to solve the gauge hierarchy problem, have been suggested. The one which 
attracted much attention being the Randall--Sundrum (RS1) 
model \cite{Randall:1999ee}, where the
 hierarchy between the electroweak (EW) and the Planck scales arises 
 from a warped higher dimensional spacetime. Variants of 
the original set-up have matured over the years. Eventually, all SM 
fields except the Higgs  (to solve the hierarchy problem, it
 is sufficient that just the Higgs --or alternative dynamics responsible 
 for electroweak symmetry breaking-- be localized at the TeV brane)
 have been promoted to bulk fields rather than brane fields. Indeed, it 
 has been realized that placing gauge fields in the bulk could lead
 to high scale unification because of the 
 logarithmic running of gauge couplings in 
 AdS$_5$ \cite{Pomarol:2000hp}. In addition, 
 allowing fermions to propagate along the extra dimension offers a 
 simple attractive mechanism for explaining the structure of Yukawa 
 couplings without introducing hierarchies at the
 level of the 5-dimensional action \cite{Grossman:1999ra}. More
  recently, it has been
  shown that EW precision constraints are much ameliorated if the 
  EW gauge symmetry is enhanced to 
  $SU(2)_L \times SU(2)_R \times U(1)_{B-L}$ \cite{Agashe:2003zs}.
  The AdS/CFT correspondence suggests that this model is dual 
  to a strongly coupled CFT Higgs sector \cite{Arkani-Hamed:2000ds}. Also, 
   the  $SU(2)_L \times SU(2)_R$
   gauge symmetry in the RS bulk implies the presence of a global custodial isospin 
symmetry of the CFT Higgs sector, thus protecting EW observables from excessive new
contributions \cite{Agashe:2003zs}. This gauge structure in warped space has also been used 
to construct higgsless models of EW symmetry breaking \cite{Csaki:2003zu}. 
Our qualitative results will apply to these models as well. 

One of the remaining phenomenological issues which has not been addressed in 
RS is the DM puzzle. No generic WIMP dark matter candidate has been identified 
yet. In this work, we would like to consider the possibility of Kaluza--Klein dark 
matter \cite{Servant:2002aq,Servant:2002hb},  so far restricted to models with flat 
TeV$^{-1}$ Universal Extra Dimensions (UED),
in warped geometries. In UED, 
%all SM fields propagate into extra dimensions and 
the Lightest Kaluza--Klein Particle (LKP) can be stable 
because of KK parity, a remnant of  translation invariance along the extra 
dimension, after the orbifold projection has been implemented. Note that for KK parity to be an 
exact symmetry, one has to assume that the boundary lagrangians at the two orbifold 
 fixed points are symmetric. A feature of models with flat toroidal TeV$^{-1}$
  extra dimensions (not necessarily UED) is the presence of a light gravitationally
   coupled radion, typically expected to have a mass in the meV range
   \cite{Chacko:2002sb}. Because  its lifetime well exceeds the
 age of the universe, it was shown in \cite{Kolb:2003mm} 
 that such light radion generically leads to cosmological 
 catastrophy like overclosure of the universe by radion oscillations. To avoid
  this problem, one has to ensure a radion stabilisation mechanism allowing 
  for a larger radion  mass.
 In contrast, this problem does not arise in RS geometry where the radion field has an EW mass 
and couples strongly \cite{Csaki:1999mp}. 
Cosmology of RS has attracted tremendous interest. In particular, it was shown
 that standard Friedmann cosmology can be recovered 
 \cite{Csaki:1999jh,Csaki:1999mp}
 and normal expansion is expected at 
 least up to a TeV temperatures above which a phase transition occurs where 
 the TeV brane is replaced by an event horizon \cite{Arkani-Hamed:2000ds,Creminelli:2001th}.
  As far as WIMP dark matter is concerned, we do
  not rely on what happens at these high temperatures since the freeze out
   temperature is typically a few tens of GeV and we can safely make a 
   standard cold dark matter relic density calculation in the RS context. 
   
   Obviously, there is no translational invariance in RS geometry, hence there is no analog of 
   KK parity conservation. Instead, the stability of a light KK mode will be related to 
    baryon number symmetry.  
   In RS, dominant baryon number violation arises through effective 
four fermion interactions localized near the TeV brane thus suppressed
 by the TeV scale only.
One solution is to localize fermions very close to the Planck brane 
where the effective cut-off is Planckian. However, it turns out that this 
suppresses too much the 4D Yukawa couplings to the Higgs on the TeV brane
and is incompatible with the spectrum of SM fermion masses.
In this paper, we impose a bulk (gauged) baryon number symmetry.
 We are also interested in
    RS GUTs. So far, such studies have focused on $SU(5)$ 
    only \cite{Goldberger:2002pc,Agashe:2002pr}. 
  We will instead assume $SO(10)$ or Pati--Salam, in which the 
   Left-Right  gauge  structure mentioned above can easily be embedded.
A priori, grand unification is at odds with imposing baryon number
 symmetry. However, baryon number symmetry can be consistent with higher dimensional
 GUT  \cite{Goldberger:2002pc,Agashe:2002pr} if the unified gauge group is broken by boundary conditions (BC)
 so that SM quarks and leptons are obtained from different bulk multiplets of the unified gauge 
  group.
%  Such procedure forbids  that TeV KK  excitations of X/Y gauge bosons couple to two SM fermions, thus avoids 
 %proton decay  coming from the s-channel exchange of TeV KK excitations of X/Y gauge bosons.
Let us start with a simple  example where $SO(10)$ is broken to the SM 
on the Planck brane by BC and the number of $\bf{16}$ representations is replicated
three times per generation:
$$
  \begin{pmatrix}\bf{u_L \ , d_L}\\u^{\prime c}_R \\ d_R^{\prime c}  \\ 
  \nu^{\prime}_L \ ,  e^{\prime}_L \\ e_R^{\prime c} \\ \nu_R^{\prime c} \\\end{pmatrix}_{B=1/3}, \ \ 
 \begin{pmatrix}u^{\prime}_L \ , d^{\prime}_L\\ \bf{u_R^c} \\ \bf{ d_R^c}  
 \\ \nu^{\prime}_L \ ,  e^{\prime}_L \\ e_R^{\prime c} \\ \nu_R^{\prime c} \\\end{pmatrix}_{-1/3}, \ \
  \begin{pmatrix}u^{\prime}_L \ , d^{\prime}_L\\u_R^{\prime c} \\ d_R^{\prime c} 
   \\ \bf{\nu_L} \ ,  \bf{e_L} \\ \bf{e_R^c} \\ \bf{\nu_R^c} \\\end{pmatrix}_{0}
$$
Only states in boldface characters have zero modes, they correspond to ($++$) BC ($+$ denotes
Neumann, $-$ Dirichlet,
first sign is for Planck brane, second for TeV brane) and are identified with the SM fermions.
Other states are ($-+$).
GUT multiplets are assigned the baryon number of the zero modes contained in them.
A $Z_3$ symmetry follows from requiring baryon number as a good quantum number:
\begin{equation}
\Phi \rightarrow e^{ 2 \pi i \left( B  - \frac{n_c - \bar{n}_c }{3} \right) } \Phi
\end{equation}
$B$ is baryon-number of $\Phi$ and  $n_c$ ($\bar{n}_c$)
is its number of color (anti-color) indices.
%$Z_3$ can be seen as a remnant discrete symmetry after the 
%unified gauge symmetry has been broken to the SM.
 Clearly, SM fields are not charged under $Z_3$.
$X$, $Y$, $X^{\prime}$, $Y^{\prime}$ and $X_s$ gauge bosons of $SO(10)$ are charged under $Z_3$
as well as lepton-like states within $\bf{16}$'s which carry baryon number and quark-like states 
 which carry non standard baryon number.
These exotic states do not have zero modes. 
Consequently, the lightest $Z_3$ charged particle 
 (LZP) cannot decay into SM particles and is stable.
Note that the baryon number gauge symmetry has to be broken since
we do not want an extra massless gauge boson in the theory.
We break it spontaneously on the Planck brane. If it is broken in such a way that 
$\Delta B \neq \frac{1}{3}, \frac{2}{3} $, we can show that proton decay is Planck suppressed \cite{RSLKP2}.
 In order to guarantee this, we impose the $Z_3$ symmetry. As a
result, the LZP is stable, and, like in supersymmetry, dark matter can arise
as a consequence of solving proton stability.  

\begin{figure}[h]
%\centerline{\includegraphics[width=1.12\hsize]{mLZP.eps}}
\centerline{\epsfxsize=.6\textwidth\epsfbox{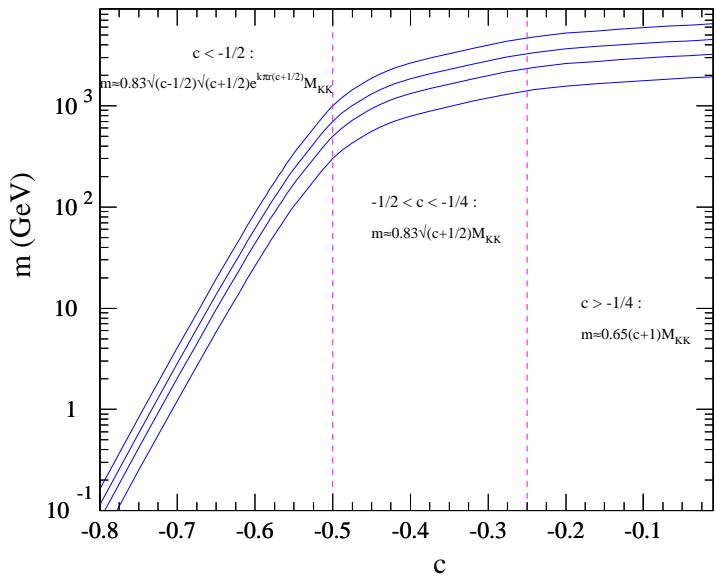}}
\caption[]{Lightest mass of ($-+$) KK fermion as a function of its $c$-parameter.
From bottom to top, $M_{KK}=$ 3, 5, 7, 10 TeV. $e^{k \pi r}\sim M_{Pl}/$TeV is the warp factor of RS geometry.}
\label{fig:LKPMASS}
\end{figure}

The question is now to identify the LZP and see whether one can naturally expect it to be neutral.
In warped space, the spectrum of fermionic KK modes is governed by two things. First, the {\it $c$-parameter} \cite{Grossman:1999ra}
which determines the localization of wave function of massless  modes along the 5th direction and therefore the size of their 4D Yukawa couplings. 
Second, it depends crucially on BC reflecting the dynamics taking place at the TeV and Planck brane. 
The  interesting thing about ($-+$) fermionic states is that they are lighter than gauge KK modes 
for  $c<1/2$ and actually much lighter for $c<0$. In the past, studies have focused on ($++$) KK fermions,
 which are always heavier than gauge KK modes, thus are unlikely to be observed
  at colliders since
  the constraints on the gauge KK mass is $M_{KK} \gtrsim 3$ TeV  \cite{Agashe:2003zs}.
Figure  \ref{fig:LKPMASS} shows the dependence on $c$  (exponential for $c<-1/2$) of the mass of  the 
first KK excitation of ($-+$) fermions. Very light ($\sim$ GeV) KK fermions are natural.
This plot is telling us that the LZP will belong to the multiplet  whose $c$ is the smallest,
namely the multiplet  with right-handed top zero
 mode which has $c$ in the range $[-1/2,0] $ to account
  for ${\cal O}(1)$ Yukawa. From now on, we will concentrate on that particular $\bf{16}$.
%%%%%%%%%%%%%%%%%%%%%%%%%%%%%%%%
%%%%%%%%%%%%%%%%%%%%%%%%%%%%%%%%%
Mass splittings between different KK states belonging to the same $\bf{16}$
 will arise from radiative corrections. 
 In addition, and maybe 
 most importantly, we expect splittings in the $c$'s due to bulk GUT breaking
 effects \cite{Agashe:2002pr}. Such effects (actually desired to 
 achieve unification through threshold type corrections) can lead to $\Delta c$'s as large as 
 $\pm 0.5$ therefore making the $c$'s of the ($-+$) states within the same {\bf 16}
almost free parameters. 
 Phenomenologically, the LZP has to be the right-handed neutrino. Indeed, it is
  well known that heavy left-handed neutrino dark matter is excluded 
  by elastic scattering experiments (unless its mass is larger than several tens of TeV)
  because of its large coupling to the $Z^0$ gauge boson, {\it e.g.}~\cite{Servant:2002hb}.
  From now on, we will therefore assume that the KK RH neutrino has the smallest c, thus is the LZP.
  In warped $SO(10)$ or Pati--Salam, the KK right-handed 
   neutrino  behaves as 
  a WIMP as follows. Its couplings to KK gauge bosons like $Z^{\prime}$, the extra $U(1)$ of $SO(10)$,  
  or $X_s$, are actually enhanced compared to SM couplings, as 
  understood from the CFT dual interpretation of KK modes as strongly
   coupled composites. However,  KK gauge bosons 
  have at least a 3 TeV mass, making cross 
  sections effectively  of the right weak scale size. It
   turns out that the LZP has actually a non-negligible
 coupling to $Z^0$ because of $Z^{\prime}$-- $Z^0$
  mixing after EW symmetry breaking as well as LZP-- $\nu^{\prime}_L$
mixing from the large Yukawa coupling between the $\bf{16}$ of $t_R$ and 
the $\bf{16}$ of ($t_L, b_L$). Such Yukawa coupling may also generate a significant LZP-Higgs coupling. 

\begin{figure}[h]
%\centerline{\includegraphics[width=1.0\hsize]{relic.eps}}
\centerline{\epsfxsize=.6\textwidth\epsfbox{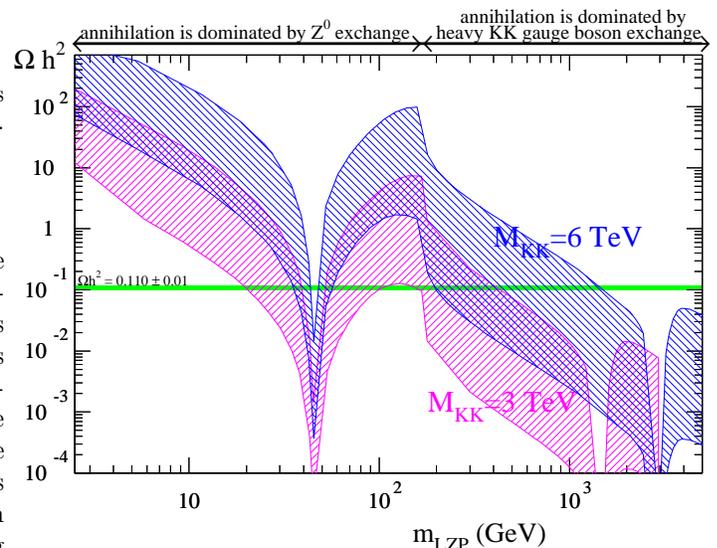}}
\caption[]{Example of relic density predictions in warped $SO(10)$ 
for two values of $M_{KK}$. $c_{t_R}=-1/2$, $c_{t_L,b_L}=0.4$, 
all  $c$'s for other fermions being larger than 1/2. 
Each region is obtained by varying both $g_{10}$ (from $g^{\prime}$ to $g_s$) 
and  $c_{\nu^{\prime}_L} \in [c_{t_L}- 0.5 , c_{t_L}+ 0.5]$.
%The excluded region corresponds to too large LZP--$Z^0$ couplings modifying the $Z^0$ width.
}
\label{fig:RELIC}
\end{figure}

We are now ready to evaluate the relic density of the LZP.
There are essentially four types of annihilation 
channels:
s-channel exchanges of $Z^0$ into any SM fermions,  
of $Z^{\prime}$ into $t \overline{t}$ , $b \overline{b}$, $ W^+ W^-$,
$Z^0$ $h$, of  Higgs
into $t \overline{t}$,  $ W^+ W^-$, $Z^0 Z^0$ and t-channel exchange of 
$X_s$ into $t_R \overline{t_R}$. Note that only fields
 localized near the TeV brane ($t$, $b$, $h$, longitudinal $W^{\pm}$ and $Z^0$)
 have large couplings to  $Z^{\prime}$ and that the only zero mode
 the LZP directly couples to is $t_R$. In a first approximation, we have not
included the Higgs exchange in our analysis. However, as explained in \cite{RSLKP2},
it becomes significant (but subdominant) only for LZP masses
between $m_t$ and $m_h$ and also dominates at the 
resonance, $m_{\mbox{\tiny{LZP}}}\sim m_h/2$, 
thus modifies the predictions of
fig. \ref{fig:RELIC} for these masses.
  There are at least 6 parameters entering the relic density 
 calculation: $c_{\mbox{\tiny{LZP}}}$ which fixes
 not only the LZP mass but also the LZP couplings,
$M_{KK}$, the gauge KK mass (mass of $Z^{\prime}$ and $X_s$)
  constrained to be $\gtrsim$ 3 TeV but 
also favoured to be as low as possible not to reintroduce a hierarchy problem,
$c_{t_R}$, $c_{b_L}$, $c_{\nu^{\prime}_L}$
 and finally $g_{10}$, the 4D $SO(10)$ gauge coupling. 
 Due to UV sensitive bulk threshold effects and finite, universal
 1--loop corrections,
 $g_{10}$ can vary from $g^{\prime}$ to $g_s$ \cite{Agashe:2002pr}.
In order to get a large 4D top Yukawa without 
pushing the 5D theory to strong coupling,
$ c_{t_R}=-1/2$ is actually favored. Our qualitative results do not depend much on the precise nature of the EW symmetry breaking sector. However, detailed quantitative predictions do. As an illustration, fig. \ref{fig:RELIC} shows 
our prediction for the relic density in the attractive case where the Higgs is a pseudo goldstone boson (PGB) \cite{Contino:2003ve} which is not exactly localized on the TeV brane but has some profile in the bulk.
For LZP masses below the top mass, annihilation is dominated by $Z^0$ 
exchange, then annihilation through $X_s$ exchange takes over until
 the LZP mass reaches the $Z^{\prime}$ pole.
 The result is that there is a large parameter space and particularly a large range of 
 LZP masses for which we can get the right relic density. 
 %%%%%%%%%%%%%%%%%%%%%%%%%%%%%%%%
%%%%%%%%%%%%%%%%%%%%%%%%%%%%%%% 

\begin{figure}[h]
%\centerline{\includegraphics[width=1.0\hsize]{elasprof3.eps}}
\centerline{\epsfxsize=.462\textwidth\epsfbox{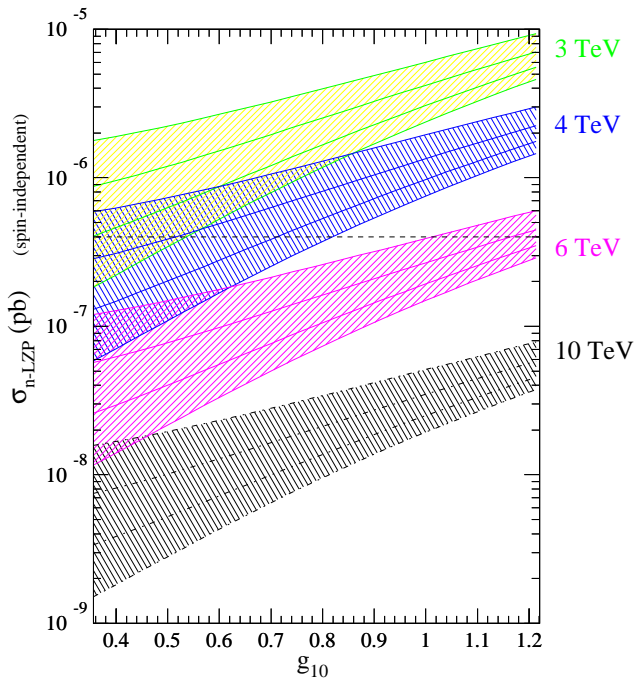}}
\caption[]{Example (where the Higgs is a PGB) of predictions for elastic scattering cross sections 
between the LZP and a nucleon (independent of LZP mass).
 For each $M_{KK} $ region, the four lines denote different values of
 the $Z^0$-LZP coupling corresponding to, from top to bottom,
$c_{\nu^{\prime}_L}=-0.1, 0.1,0.4,0.9$.
Horizontal dotted line indicates present experimental limit, which only applies for some range of WIMP masses, see \cite{DMplotter} for instance.  For this range of LZP masses, only $g_{10}$ values below 0.55 survive in the $M_{KK}= 3$ TeV case.}
\label{fig:DIRECT}
\end{figure}

Our LZP being a Dirac particle, with significant coupling to the $Z^0$, we predict
 large cross sections for its elastic scattering off nuclei (the calculation
is similar to the one in \cite{Servant:2002hb}).  Comparatively,
scattering via Higgs exchange is always negligible.
As shown in Fig \ref{fig:DIRECT},
 the entire parameter space will be tested at near future direct detection experiments. As an
  illustration, we show the $M_{KK}=10$ TeV case,  disfavoured as far as
  fine-tuning of the Higgs mass is concerned. Even this extreme case, in which  
  discovery of KK modes at colliders would be hopeless, could easily be probed
by direct DM searches.
%%%%%%%%%%%%%%%%%%%%%%%%%%%%%%%%
%%%%%%%%%%%%%%%%%%%%%%%%%%%%%%%
%%%%%%%%%%%%%%%%%%%%%%%%%%%%%%%%
\begin{figure}[h]
\centerline{\epsfxsize=.4\textwidth\epsfbox{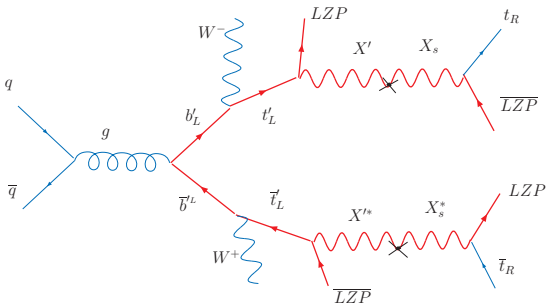}}
%\centerline{\includegraphics[width=1.0\hsize]{collider_BL.eps}}
\caption[]{Pair production and decay of $b^{\prime}_L$, GUT partner of the LZP. Decay
 occurs through $X^{\prime}$--$X_s$ mixing due to bulk $SO(10)$ breaking.}
\label{fig:COLLIDER}
\end{figure}
%%%%%%%%%%%%%%%%%%%%%%%%%%%%%%
Pair production of our WIMP at future accelerators
can be observed only for the largest values of the $Z^0$--LZP coupling, 
which are already ruled out by DM direct searches. 
However, there are very promising collider signatures associated with 
the pair production of the next  lightest exotic KK modes, in the
 same {\bf 16} as the LZP. Those are also expected to have 
 $c\sim -1/2$ with masses in the few hundred GeV--1 TeV range. Such KK 
 modes cannot decay easily.  Details will be presented in
\cite{RSLKP2}.  As an example, we show in Fig.\ref{fig:COLLIDER} the
  decay of  $b_L^{\prime}$, which has to go through 
  a 4-body decay: two LZPs, a top and a $W$, leading to quite a
  unique signature.
   In a significant part of parameter space, such 4-body decay
 will be forbidden kinematically and $b_L^{\prime}$ may lead to a ionisation
  track in the detector, something also easy to search for.
   
   In summary, we showed that solving the problem of baryon-number violation in
higher-dimensional warped GUT by imposing a discrete $Z_3$ symmetry leads to the stability
 of a light KK fermion, which acts as a viable dark matter candidate.
 We also emphasized the interesting phenomenology associated with
KK fermions with ($-+$) type of  boundary conditions in warped geometry, our DM candidate being one of them. 
It is expected that all fields within a multiplet may not
 have the same BC, in particular 
in GUT theories where the gauge symmetry is broken by BC. 
 We predict the KK modes in the GUT multiplet whose zero mode is $t_R$ to be 
 light ($\lesssim$ TeV) and observable at future colliders.
 Model building issues, further phenomenological aspects and 
 technical details of these models will be provided elsewhere\cite{RSLKP2}.  
\vspace{-1.cm} 
%%%%%%%%%%%%%%%%%%%%%%%%%% 
\section*{Acknowledgments} 
\vspace{-.3cm} 
This work benefitted from diverse discussions with 
C. Balazs, M. Battaglia, A. Birkedal, C. Cs\'aki,  H. Frisch, C. Grojean,
I. Hinchliffe, D. Morrissey, R. Sundrum, T. Tait and N. Weiner.
We thank the hospitality of the Aspen Center for 
Physics where this work was initiated. 
G.S also thanks KITP Santa 
Barbara and University of Michigan.
% where part of this work was carried out.
K.~A.~is supported by the Leon Madansky fellowship
and NSF Grant P420D3620414350.
G.~S.~is supported in part by the US 
Department of Energy, High Energy Physics Division, under contract 
W-31-109-Eng-38 and also by the David and Lucile Packard Foundation.  
%%%%%%%%%%%%%%%%%%%%%%%%%%%%%%%%%%%%%%%%% 
%%%%%%%%%%%%%%%%%%%%%%%%%%%%%%%%%%%%%%%%% 

\end{document}